# MILLIMETER AND SUBMILLIMETER SPECTROSCOPY OF THE DEUTERATED MOLECULAR ION SD$^+$

Mitsunori Araki, Valerio Lattanzi, Christian P. Endres, and Paola Caselli
Center for Astrochemical Studies, Max-Planck-Institut für extraterrestrische Physik, Giessenbachstrasse 1, Garching bei München, 85748, Germany, araki@mpe.mpg.de

## ABSTRACT

Seven rotational and fine-structure transitions of the deuterated molecular ion SD$^+$ in the $X\,^3\Sigma^-$ ground electronic state have been measured in the 271–863 GHz region in the laboratory. This ion has been produced by DC-glow discharge using a mixture of D$_2$S and argon in a free space cell in a temperature range of −140 to −160°C. The rotational, centrifugal distortion, spin-spin interaction, and hyperfine constants have been determined; the standard deviation of the residuals in the fitting is 109 kHz. The set of obtained spectroscopic parameters provides a list of accurate sub-millimeter rest frequencies of SD$^+$ for astronomical detection. We have investigated lines of SD$^+$ toward the quasar PKS 1830-211 using the ALMA archive, as the $z = 0.89$ molecular absorber exists in front of this quasar. A data set covering the 297 GHz region includes the $N_J = 2_3$–$1_2$ transition at 561 GHz due to redshift, providing an upper limit of the column density $N_{\text{tot}} = 3 \times 10^{12}$ cm$^{-2}$ for SD$^+$.

*Subject Keywords*: Astrochemistry—ISM: molecules—methods: laboratory—molecular data

## 1. Introduction

The isotopic ratio of a molecule provides information on the chemical evolution in space. Although the hydrogen isotopic ratio in space is D/H = $(1.52 \pm 0.08) \times 10^{-5}$ (Linsky 2003), this ratio is one order of magnitude larger on Earth oceans, i.e., $1.56 \times 10^{-4}$ (Berglund & Wieser 2011). This has probably its origins in the deuterium enrichment by the exothermic reaction of H$_3^+$ + HD → H$_2$D$^+$ + H$_2$ (Millar et al. 1989) in a low-temperature interstellar cloud (e.g. Cleeves et al. 2014). Hence, tracing an isotopic ratio of hydrogen in a molecule provides clues on the chemical evolution in space (Caselli & Ceccarelli 2012, Ceccarelli et al. 2014). In particular, the ratio of a protonated atom with its deuterated species can trace the initial phase of chemical evolution due to the contribution of the reaction pair of H$_3^+$ + A → AH$^+$ + H$_2$ and H$_2$D$^+$ + A → AD$^+$ + H$_2$.

At present, 295 species of interstellar molecules have been found in space (Araki, McGuire 2022). Several protonated atoms, CH$^+$ (Douglas & Herzberg 1941), OH$^+$ (Wyrowski et al. 2010), SH$^+$ (W3 IRS5, Benz et al. 2010), ClH$^+$ (De Luca et al. 2012), HeH$^+$ (Güsten et al. 2019), and ArH$^+$ (Barlow et al. 2013) have been detected so far. Following the first detection of SH$^+$, this ion was observed in the clouds toward Sagittarius B2 (Menten et al. 2010), the Orion Bar (Müller et al. 2014), and the z = 0.89 molecular absorber in front of the quasar PKS 1830-211 (Muller et al. 2014). On the contrary, the detection of deuteronated species, having a deuteron D$^+$, is limited to CD$^+$ (Möller et al. 2021). The difficulty of detection for OD$^+$, ClD$^+$, and HeD$^+$ comes from the fact that these light ions produce a limited number of rotational lines in the shorter-submillimeter region up to 1 THz because of their large rotational constants. On the other hand, SD$^+$ and ArD$^+$ can produce a sufficient number of lines in the millimeter region due to their smaller rotational constants. In particular, the richness of sulfur-bearing species in molecular clouds gives us a chance to detect SD$^+$ using a radio telescope. However, rest frequencies of rotational and fine-structure transitions for SD$^+$ were not measured yet in a laboratory because of the difficulty in producing the molecular ion in an experimental system.

Spectroscopy of deuterated species of these protonated atoms, i.e., CD$^+$ (Amano 2010), OD$^+$ (Verhoeve et al. 1986), HeD$^+$ (Matsushima et al. 1997), and ArD$^+$ (Bowman et al. 1983), have been reported thanks to the measurements of their rotational transitions in laboratories. These data allow us to detect these ions in space by radio observations. Spectroscopy of the rotational and fine-structure transitions of $SH^+$ has been reported by laboratory measurements (Hovde & Saykally 1987, Savage et al. 2004, Brown & Müller 2009, Halfen & Ziurys 2015) and astronomical observations (Müller et al. 2014). However, for $SD^+$, spectroscopy is limited to the electronic transition of $A^3\Pi - X^3\Sigma^-$ (Rostas et al. 1984) and the vibrational transition of $v = 1-0$ (Zeitz et al. 1987), i.e., its rotational transitions have not yet been measured.

In this paper, we report the laboratory measurements of the rotational and fine-structure transitions for $SD^+$ in the $X^3\Sigma^-$ ground electronic state by millimeter and submillimeter spectroscopy. The frequencies obtained by our measurements can be used as the rest frequencies to detect this ion in space. Additionally, we also report an upper limit of $SD^+$ column density toward a quasar using archival ALMA data.

**2. Experimental Details**

The rotational and fine-structure transitions of $SD^+$ were observed using the combination of the Center for Astrochemical Studies Absorption Cell (CASAC) and the spectroscopy instrument consisting of a synthesizer and an active multiplier chain as described below (Bizzocchi et al. 2017). This combination was employed to record the spectrum in the 271–863 GHz range. The spectroscopic instrument uses a synthesizer (Agilent Technologies E8257D) as a primary source in the 9.2–13.9 GHz band and an active multiplier chain (Virginia Diodes Inc.) as a source of millimeter radiation in the 82.5–125 GHz band. The higher frequency output is achieved by using cascaded multiplier stages. A closed-cycle He-cooled InSb hot electron bolometer operating at 4 K is used as a detector. The 2f-frequency-modulation technique is employed to improve the signal-to-noise ratio (S/N), where the source is sinewave modulated at 15 kHz and the detector output is demodulated at twice this frequency by a lock-in amplifier. Therefore, the second derivative of the actual absorption profile is recorded by the acquisition system. In the present measurements, the modulation depth was set at 500 kHz in the case of the 560 GHz region, and it was tuned according to frequency.

The present cell is a free-space discharge type with a length of 3 m and an inner diameter of 5 cm. A pair of 25-cm-length stainless-steel cylindrical electrodes having a distance of 1.3 m from each other was installed in the cell. A 1.8-m region covering the two electrodes and a discharge space in between was cooled by circulating liquid nitrogen through a Teflon tube coil around the glass cell. $SD^+$ was generated by the DC-glow discharge using a mixture of $D_2S$ and argon with a ratio of 5:100 in the cell under a temperature range of $-140$ to $-160°C$, where the flow rate of argon was set at 10 sccm (standard cc/min). To achieve a pressure of 90 mTorr in the cell, the pumping speed was controlled by a butterfly valve. The current and voltage of discharge were set at 200 mA and around 900 V, respectively, which were optimized by monitoring the $SH^+$ line. The observed $N_J = 2_3–1_2$ transition of $SD^+$ is shown in Figure 1.

Since the present system uses DC discharge, a Doppler shift of the absorption frequency by the drift of the ion itself was expected. The $SH^+$ line at 526048 MHz was used to test the shift. However, compared with the line frequency observed in AC discharge (Halfen & Ziurys 2015), a significant frequency shift in the DC discharge was not observed. Hence, we presume that the Doppler shift caused by the drift velocity of $SD^+$ is negligible in the present measurements.

**3. Results and Discussion**
*3.1. Line Analysis*

Firstly, the $N_J = 2_2–1_1$ and $3_3–2_2$ transitions between the $N = J$ rotational levels have been searched because the line frequencies of these transitions are determined by only the rotational constant $B_0$, except for small contributions of the centrifugal distortion constant $D_0$. The lines of both transitions were found at the estimated frequencies using $B_0$ reported by Zeitz et al. (1987), i.e., at 567431 and 850911 MHz, respectively. Secondly, the $N_J = 2_3–1_2$ transition between the $N \neq J$ rotational levels has been searched at $560848 \pm 7$ MHz, which was estimated by the set of molecular constants by Zeitz et al. (1987). However, no line of this transition was found in this spectral region. Hence, we expected a deviation of the spin-spin constant $\lambda_0$. This is because this constant determined by Zeitz et al. (1987) is not the same as that by Rostas et al. (1984), although $B_0$ and the spin-rotation constant $\gamma_0$ agree well between both reports, as listed in Table 1. This deviation of $\lambda_0$ would come from the difficulty of determining $\lambda_0$ by using the rotational-vibrational and rotational-electronic transitions. By searching the $\pm 50$ MHz region, one line having similar physical and chemical behavior, i.e., dependence on temperature of the cell and a mixing ratio of $D_2S$, as the $N_J = 2_2–1_1$ and $3_3–2_2$ lines was found at 560807.6 MHz. This line constrains $\lambda_0$ and subsequently the frequencies of the other transitions between the $N \neq J$ rotational levels were newly estimated. The lines of these transitions were found at the estimated frequencies. Hence, the assignment of the $N_J = 3_3–2_2$ transition was confirmed. Finally, the seven rotational and fine-structure transitions were measured with an experimental accuracy of 100 kHz, determined from their line widths, S/N, and baseline. The lines are listed in Table 2.

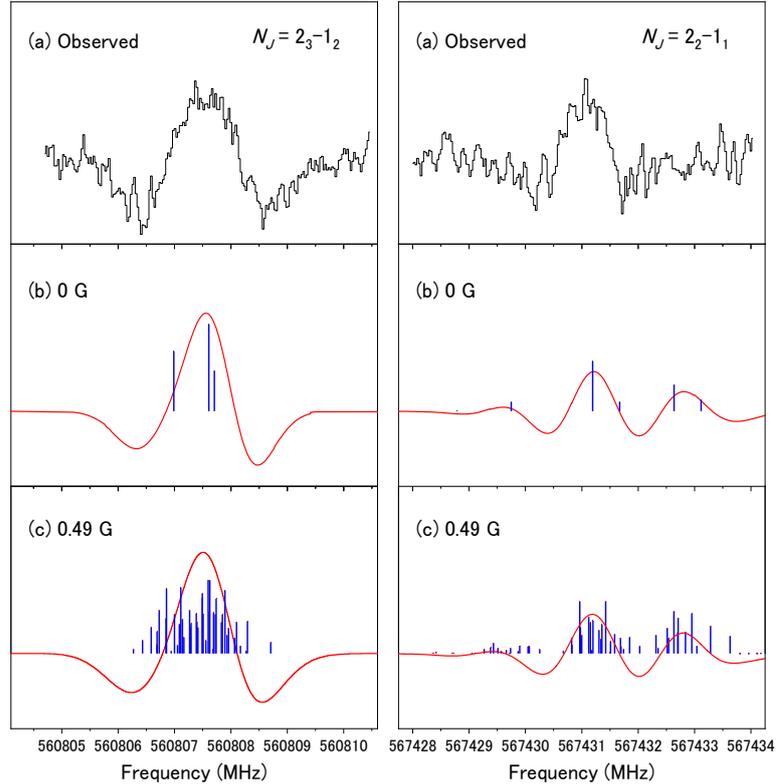

**Figure 1.** (a) Observed line profiles of the $N_J = 2_3–1_2$ and $2_2–1_1$ transitions for SD$^+$. Seven and five spectra in the same regions are averaged, and then their total integration times reach 5144 and 1823s, respectively, with a 30-ms time constant. (b) The blue vertical segments show the hyperfine structure under a non-magnetic field. (c) The blue vertical segments describe the effect of the Earth's magnetic field of 0.49 G in Munich (International Real-time Magnetic Observatory Network), where the $\Delta M = 0$ components are half-weighted. The segmented spectra were simulated with PGOPHER (Western). The red lines in (b) and (c) are Gaussian profiles having a 0.5 MHz width in the 2f-frequency modulation. The relative intensities between the left- and right-side drawings for (b) and (c) reflect the temperature at which the experiment has taken place, i.e. 130 K, but those between (b) and (c) and between the profiles and the vertical segments are arbitrary. The weak line feature at around 567433 MHz for $N_J = $

To determine the hyperfine constants $b_F(D)$ and $c(D)$, the initial values of both constants were estimated from those of SH$^+$ (Halfen & Ziurys 2015) scaled by the ratio between the analogous constants of OH$^+$ (Markus et al. 2016) and OD$^+$ (Verhoeve et al. 1986), because no values of $b_F(D)$ and $c(D)$ were reported so far. Fitting was executed by using the program SPFIT (Pickett 1991) and uncertainties of obtained spectroscopic parameters were evaluated by using the program PIFORM (Kisiel 2022). The spin-rotation constant $\gamma_0$ was fixed to the value reported by Zeitz et al. (1987) because this constant has a large correlation with $\lambda_0$ and $b_F(D)$. The rotational, centrifugal distortion, spin-spin interaction, and hyperfine constants of this ion were precisely determined (Table 1). The precision of the rotational constant $B_0$ has been enhanced by nearly one order of magnitude compared to that of Zeitz et al. (1987). The rms value of 109 kHz was achieved in this fitting.

All of the lines were measured under the Earth's magnetic field of 0.49 G in Munich (International Real-time Magnetic Observatory Network), as also done for the rotational transitions of OH$^+$ (Bekooy et al. 1985) and OD$^+$ (Verhoeve et al. 1986) in Nijmegen. The Zeeman splitting for these lines was simulated by using PGOPHER (Western) with the g-factors ($g_1 = 0.0055$, $g_s = 2.0023$) reported by Zeitz et al. (1987) assuming our local magnetic field. As examples of the splitting, the cases of the $N_J = 2_3–1_2$ and $2_2–1_{11}$ transitions are shown in Figures 1b and 1c under the 0 and 0.49 G magnetic fields, respectively. This simulation suggests that the effect of the Earth's magnetic field is negligible on the widths and frequencies of our lines when using a 500-kHz modulation depth and a 90-mTorr gas pressure. Therefore, the frequencies measured under the Earth's magnetic field will have the same experimental errors as those under no magnetic field. This is supported by the fact that the SH$^+$ line at 526048 MHz was measured in our setup, providing the same frequency as that by the μ-metal covered cell within the measurement accuracies.

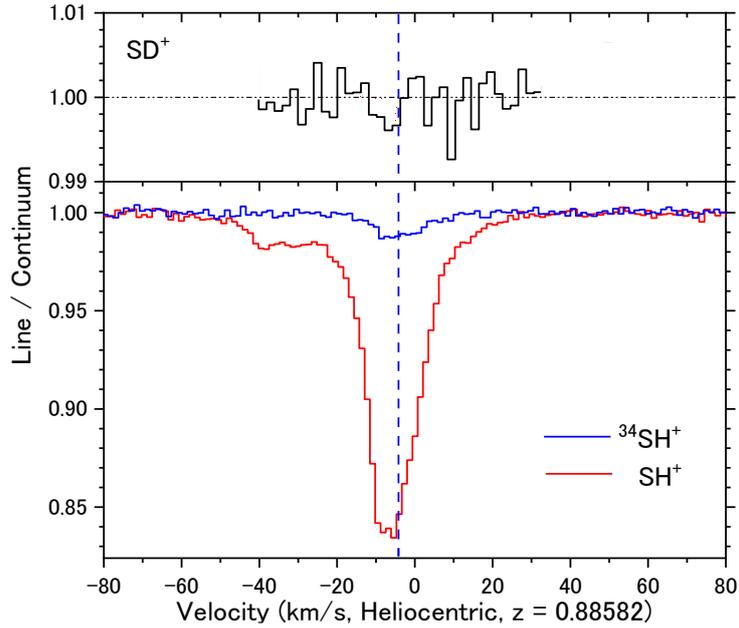

**Figure 2.** Absorption lines in the SW component of the molecular absorber toward the quasar PKS 1830-211 for SD$^+$, SH$^+$, and $^{34}$SH$^+$. The upper histogram is the $N_J = 2_3–1_2$ transition of SD$^+$ sampled from the 0.25-arcsec-diameter circular area having the center of [α(J2000), δ(J2000)] = (8$^h$33$^m$39$^s$.89, −21°03′,40″.55) in ALMA 2013.1.00020.S data set (PI: S. Muller) by using CASA 6.5.3, and the lower histograms are the $N_J = 1_2–0_1$ transition of SH$^+$ and $^{34}$SH$^+$ sampled from the 0.04-arcsec-diameter circular area in ALMA 2018.1.00051.S data set (PI: S. Muller). The size of the circular area was set in a viewer window of CASA to cover emission from the SW component in each data set. These spectra were obtained selecting a region encompassing the emission from the SW component in the "fits" data stored in the ALMA achieve. The vertical dashed line shows the velocity of ND reported by Muller et al. (2020).

The determined molecular constants given in Table 1 were adopted to predict the frequencies and their uncertainties (1σ) for the rotational and fine-structure transitions of SD$^+$ in the $X^3\Sigma^−$ ground electronic state by using the program SPCAT (Pickett 1991). The predicted uncertainties fall within the range of 0.11–2.86 MHz for transitions up to 1 THz. This precision is enough to permit an identification of this ion in almost every case of an astronomically obtained spectrum. The predicted rest frequencies are listed in Table 3. The Einstein A coefficient, the unitless line strengths $S$ (Mangum & Shirley 2015), and the energy of the lower level of the transitions ($E_{low}$) are computed and given in the same table. The listed parameters in Table 3 can be used to evaluate the intensities of individual transitions, and finally achieve a definitive identification of SD$^+$ in space.

### 3.2. Comparison with the Current Astronomical Data

We have searched for lines of SD$^+$ in the ALMA archive. The z = 0.89 molecular absorber, a nearly face-on spiral galaxy (Winn et al. 2002), is located in front of the quasar PKS 1830-211. SH$^+$ was detected at the SW component of the molecular absorber (Muller et al. 2017). Although this molecular ion has been detected in other clouds, i.e., W3 IRS5 (Benz et al. 2010), Sagittarius B2 (Menten et al. 2010), and the Orion Bar (Müller et al. 2014), only this SW component shows abundant SH$^+$ with a simple velocity structure, as shown in Figure 2. A cube data set covering the 297 GHz region is available, and this includes the $N_J = 2_3–1_2$ transition of SD$^+$ at 561 GHz redshifted to this region (using the formula $v(z) = v_0/(1 + z)$, with $z$ = 0.88582). In the SW component, a weak line having

- 4 -

an S/N of 2 is located at the same velocity with the peaks of SH$^+$ and $^{34}$SH$^+$ (Figure 2). To derive the column density of SD$^+$, we used the coefficient $\alpha$ = $1.23 \times 10^{14}$ cm$^{-2}$ km$^{-1}$ s assuming a dipole moment of 1.087D (Senekowitsch et al. 1985) and a rotational temperature of 5.14 K. This temperature is the cosmic microwave background temperature at z = 0.89 and was used to derive the column density of SH$^+$ (Muller et al. 2017). This coefficient $\alpha$ gives a column density, $N_{col}$, from the integrated opacity, $\int \tau\, dv$, according to the equation $N_{col} = \alpha \int \tau\, dv$ (Muller et al. 2014). A continuum covering factor is assumed to be $f_c$ = 1 (Eq. (1) in Muller et al. 2014), i.e., all background flux is intercepted by the absorber. The three hyperfine components $F$ = 3–2, 4–3, and 2–1 having line strengths of $S$ = 2.52, 3.65, and 1.70, respectively, are superimposed in this line. The column density **upper limit** is derived to be < $3 \times 10^{12}$ cm$^{-2}$ from the observed line feature.

This column density is comparable with those of ND ($1.7 \times 10^{12}$ cm$^{-2}$), o-NH$_2$D ($0.6 \times 10^{12}$ cm$^{-2}$), and HDO ($1.0 \times 10^{12}$ cm$^{-2}$) observed in the same molecular absorber (Muller et al. 2020). However, the column density of SH$^+$ was reported to be $3.9 \times 10^{13}$ cm$^{-2}$ (Muller et al. 2017). The column densities of SD$^+$ and SH$^+$ would then provide an abundance ratio of SD$^+$/SH$^+$ < 7%, which is significantly larger than the known isotopic ratios of ND/NH (0.07–0.7%), NH$_2$D/NH$_3$ (0.01–0.3%), and HDO/H$_2$O (~0.1%) (Muller et al. 2020). To detect this deuterated ion in this source, the transitions from the lowest rotational level $N_J = 0_1$ need to be observed, as they are more likely to be detected because of the low excitation temperature measured for other molecules toward this absorber.

### 4. Summary

To detect an interstellar molecule in space, the laboratory production of the molecule, followed by measurements of precise rest frequencies, is essential. We have measured the seven rotational and fine-structure transitions of SD$^+$ in the laboratory in the 271–863 GHz frequency range. The molecular constants have been determined with a precision of ≤ 1 MHz. Especially, the spin-spin constant $\lambda_0$ has been modified from the values determined by the electronic transition and the vibrational transition reported so far. These newly determined constants are sufficient to derive the rest frequencies in the submillimeter region to detect this molecular ion in space.

ALMA archive data was used to seek lines of SD$^+$ in space. A tentative absorption at the redshifted frequency of the $N_J = 3_2$–$2_1$ transition was found in a molecular absorber toward PKS 1830-211. The intensity of this peak provides an upper limit of the SD$^+$ column density. The derived upper limit of the isotopic ratio SD$^+$/SH$^+$ is significantly larger than that measured in other molecules so more sensitive observations are needed to detect SD$^+$. A lower excitation line ($N_J = 1_2$–$0_1$ or $1_1$–$0_1$) would allow us to better constrain the SD$^+$/SH$^+$ ratio.

The authors acknowledge the financial support of the Max Planck Society. This paper makes use of the following ALMA data: ADS/JAO.ALMA#2013.1.00020.S, ADS/JAO.ALMA#2018.1.00051.S. ALMA is a partnership of ESO (representing its member states), NSF (USA), and NINS (Japan), together with NRC (Canada), MOST and ASIAA (Taiwan), and KASI (Republic of Korea), in cooperation with the Republic of Chile. The Joint ALMA Observatory is operated by ESO, AUI/NRAO, and NAOJ. M.A. thanks Yamada Science Foundation and ESPEC Foundation for Global Environment Research and Technology.

**Table 1**
Molecular Constants of SD$^+$ in MHz [a]

|  | This work [b] | Zeitz et al. (1987) | Rostas et al. (1984) |
|---|---|---|---|
| $B_0$ | 141889.428(83) | 141888.77(45) | 141891.2(48) |
| $D_0$ | 3.9387(49) | 3.8646 [c] | 3.8646(90) |
| $\lambda_0$ | 167472.6(10) | 171254.9(95) | 172740(150) |
| $\gamma_0$ | −2560.2 [d] | −2560.2(62) | −2558.4(90) |
| $b_F(D)$ | −7.17(69) |  |  |
| $c(D)$ | 15.1(18) |  |  |

[a] Values in parentheses denote the uncertainties (1σ) and apply to the last digit of the values.
[b] The rms of the fitting is 109 kHz.
[c] Fixed at the value of Rostas et al.
[d] Fixed at the value of Zeitz et al.

**Table 2**
Measured Frequencies of Rotational and Fine-Structure Transitions for SD$^+$ in MHz

| $N'–N''$ | $J'–J''$ | $F'–F''$ | $\nu_{obs}$ | $\nu_{obs} - \nu_{calc}$ |
|---|---|---|---|---|
| 1–0 | 2–1 | 3–2 | 271406.064 | 0.022 |
|  | 1–1 | 2–2 | 429975.446 | 0.001 |
| 2–1 | 3–2 | 3–2 |  | 0.594 |
|  |  | 4–3 [a] | 560807.581 | −0.028 |
|  |  | 2–1 [a] |  | −0.125 |
|  | 2–1 | 3–2 | 567431.024 | −0.168 |
| 3–2 | 4–3 | 4–3 [a] |  | 0.284 |
|  |  | 5–4 [a] | 846114.509 | −0.047 |
|  |  | 3–2 [a] |  | −0.133 |
|  | 3–2 | 4–3 | 850910.932 | −0.018 |
|  |  | 3–2 | 850911.375 | −0.056 |
|  |  | 2–1 |  | −0.296 |
|  | 2–1 | 3–2 | 863267.902 | 0.134 |

[a] Blended. In the fitting, each transition was weighted depending on its relative intensity.

**Table 3**
Calculated Rest Frequencies of Rotational and Fine-Structure Transitions for SD$^+$

| N′ − N″ | J′ − J″ | F′ − F″ | ν (MHz) | uncertainty[a] (MHz) | $E_{\text{low}}$ (K) | A (s$^{-1}$) as log A | S[b,c] |
|---|---|---|---|---|---|---|---|
| 1 − 0 | 0 − 1 | 1 − 0 | 97565.29 | 0.87 | 0.0 | −5.64667 | 0.18 |
|  |  | 1 − 1 | 97573.22 | 0.62 | 0.0 | −5.16933 | 0.53 |
|  |  | 1 − 2 | 97589.08 | 1.73 | 0.0 | −4.94726 | 0.88 |
|  | 2 − 1 | 2 − 1 | 271404.41 | 0.29 | 0.0 | −4.14100 | 1.31 |
|  |  | 1 − 0 | 271405.97 | 0.27 | 0.0 | −4.49320 | 0.58 |
|  |  | 3 − 2 | 271406.04 | 0.13 | 0.0 | −3.86990 | 2.45 |
|  |  | 1 − 1 | 271413.90 | 0.67 | 0.0 | −4.61809 | 0.44 |
|  |  | 2 − 2 | 271420.28 | 1.24 | 0.0 | −4.61798 | 0.44 |
|  |  | 1 − 2 | 271429.77 | 2.13 | 0.0 | −5.79406 | 0.03 |
|  | 1 − 1 | 1 − 0 | 429948.76 | 2.60 | 0.0 | −4.06730 | 0.24 |
|  |  | 0 − 1 | 429955.25 | 2.27 | 0.0 | −4.06739 | 0.24 |
|  |  | 1 − 1 | 429956.69 | 1.93 | 0.0 | −4.19229 | 0.18 |
|  |  | 2 − 1 | 429959.58 | 1.50 | 0.0 | −3.97039 | 0.29 |
|  |  | 1 − 2 | 429972.56 | 0.98 | 0.0 | −3.97047 | 0.29 |
|  |  | 2 − 2 | 429975.45 | 0.18 | 0.0 | −3.49327 | 0.88 |
| 2 − 1 | 1 − 1 | 0 − 1 | 381232.42 | 0.75 | 20.6 | −4.17193 | 0.26 |
|  |  | 1 − 2 | 381238.91 | 0.43 | 20.6 | −4.07502 | 0.33 |
|  |  | 1 − 1 | 381241.79 | 0.65 | 20.6 | −4.29682 | 0.20 |
|  |  | 1 − 0 | 381243.23 | 1.11 | 20.6 | −4.17192 | 0.26 |
|  |  | 2 − 2 | 381257.66 | 1.75 | 20.6 | −3.59780 | 0.99 |
|  |  | 2 − 1 | 381260.54 | 2.39 | 20.6 | −4.07489 | 0.33 |
|  | 3 − 2 | 3 − 2 | 560806.99 | 0.16 | 13.0 | −3.05874 | 2.52 |
|  |  | 4 − 3 | 560807.61 | 0.12 | 13.0 | −2.89844 | 3.65 |
|  |  | 2 − 1 | 560807.71 | 0.19 | 13.0 | −3.22944 | 1.70 |
|  |  | 2 − 2 | 560817.20 | 0.94 | 13.0 | −3.96183 | 0.32 |
|  |  | 3 − 3 | 560821.22 | 1.30 | 13.0 | −3.96183 | 0.32 |
|  | 2 − 1 | 1 − 2 | 567428.79 | 0.85 | 20.6 | −4.90097 | 0.02 |
|  |  | 2 − 2 | 567429.75 | 0.54 | 20.6 | −3.72487 | 0.37 |
|  |  | 3 − 2 | 567431.19 | 0.14 | 20.6 | −2.97657 | 2.10 |
|  |  | 1 − 1 | 567431.67 | 0.18 | 20.6 | −3.72487 | 0.37 |
|  |  | 2 − 1 | 567432.63 | 0.46 | 20.6 | −3.24767 | 1.13 |
|  |  | 1 − 0 | 567433.12 | 0.62 | 20.6 | −3.59987 | 0.50 |
|  | 1 − 0 | 0 − 1 | 713615.89 | 2.86 | 4.7 | −3.58327 | 0.16 |
|  |  | 1 − 1 | 713625.27 | 1.90 | 4.7 | −3.10616 | 0.47 |
|  |  | 2 − 1 | 713644.02 | 0.18 | 4.7 | −2.88435 | 0.78 |
|  | 2 − 2 | 1 − 1 | 725974.46 | 2.45 | 13.0 | −3.72016 | 0.18 |
|  |  | 2 − 1 | 725975.42 | 2.28 | 13.0 | −4.19726 | 0.06 |
|  |  | 1 − 2 | 725983.95 | 1.64 | 13.0 | −4.19726 | 0.06 |
|  |  | 2 − 2 | 725984.92 | 1.43 | 13.0 | −3.53165 | 0.28 |
|  |  | 3 − 2 | 725986.36 | 1.20 | 13.0 | −4.18145 | 0.06 |
|  |  | 2 − 3 | 725999.15 | 0.52 | 13.0 | −4.18145 | 0.06 |
|  |  | 3 − 3 | 726000.59 | 0.23 | 13.0 | −3.27835 | 0.50 |
| 3 − 2 | 2 − 2 | 1 − 2 | 677069.54 | 0.62 | 47.9 | −4.25798 | 0.06 |

| | | | | | | | | | | |
|---|---|---|---|---|---|---|---|---|---|---|
| | | | | 1 | – | 1 | 677070.51 | 0.66 | 47.9 | −3.78088 | 0.19 |
| | | | | 2 | – | 3 | 677078.55 | 0.33 | 47.9 | −4.24217 | 0.07 |
| | | | | 2 | – | 2 | 677080.00 | 0.61 | 47.9 | −3.59237 | 0.30 |
| | | | | 2 | – | 1 | 677080.96 | 0.89 | 47.9 | −4.25797 | 0.06 |
| | | | | 3 | – | 3 | 677094.23 | 1.80 | 47.9 | −3.33906 | 0.54 |
| | | | | 3 | – | 2 | 677095.68 | 2.08 | 47.9 | −4.24216 | 0.07 |
| | 4 | – | 3 | 4 | – | 3 | 846114.23 | 0.13 | 39.9 | −2.47308 | 3.63 |
| | | | | 5 | – | 4 | 846114.56 | 0.12 | 39.9 | −2.35798 | 4.74 |
| | | | | 3 | – | 2 | 846114.64 | 0.11 | 39.9 | −2.59118 | 2.77 |
| | | | | 3 | – | 3 | 846124.85 | 0.90 | 39.9 | −3.64917 | 0.24 |
| | | | | 4 | – | 4 | 846127.84 | 1.18 | 39.9 | −3.64917 | 0.24 |
| | 3 | – | 2 | 3 | – | 3 | 850909.99 | 0.41 | 47.9 | −3.44534 | 0.30 |
| | | | | 2 | – | 2 | 850910.71 | 0.19 | 47.9 | −3.44534 | 0.30 |
| | | | | 4 | – | 3 | 850910.95 | 0.12 | 47.9 | −2.38194 | 3.43 |
| | | | | 3 | – | 2 | 850911.43 | 0.12 | 47.9 | −2.54224 | 2.37 |
| | | | | 2 | – | 1 | 850911.67 | 0.18 | 47.9 | −2.71294 | 1.60 |
| | 2 | – | 1 | 1 | – | 2 | 863241.64 | 2.49 | 38.9 | −4.29248 | 0.03 |
| | | | | 2 | – | 2 | 863252.09 | 1.46 | 38.9 | −3.11637 | 0.43 |
| | | | | 1 | – | 1 | 863260.39 | 0.59 | 38.9 | −3.11637 | 0.43 |
| | | | | 3 | – | 2 | 863267.77 | 0.14 | 38.9 | −2.36826 | 2.42 |
| | | | | 1 | – | 0 | 863269.76 | 0.44 | 38.9 | −2.99147 | 0.58 |
| | | | | 2 | – | 1 | 863270.84 | 0.47 | 38.9 | −2.63926 | 1.30 |
| 4 – 3 | 3 | – | 3 | 2 | – | 3 | 966889.69 | 0.91 | 88.7 | −4.37550 | 0.02 |
| | | | | 2 | – | 2 | 966890.41 | 0.84 | 88.7 | −3.47240 | 0.19 |
| | | | | 3 | – | 4 | 966899.66 | 0.43 | 88.7 | −4.37009 | 0.02 |
| | | | | 3 | – | 3 | 966900.62 | 0.46 | 88.7 | −3.35069 | 0.25 |
| | | | | 3 | – | 2 | 966901.34 | 0.61 | 88.7 | −4.37549 | 0.02 |
| | | | | 4 | – | 4 | 966914.23 | 1.55 | 88.7 | −3.19399 | 0.36 |
| | | | | 4 | – | 3 | 966915.19 | 1.71 | 88.7 | −4.37008 | 0.02 |

[a] 1σ  
[b] Only the lines having a line strength $S$ larger than 0.01 are listed in this table.  
[c] unitless